\documentclass[a4paper,10pt]{ica2013_2}

\usepackage[utf8x]{inputenc}
\usepackage{xcolor}
\usepackage{amsmath,epsfig,amsfonts,amssymb,epsfig,amsbsy}
\usepackage{mathrsfs}
\usepackage{wrapfig}
\usepackage{titlesec}

\newcommand{\Real}{\mathbb{R}}

\def \bP{\mathbf{P}}

\def \bX{\mathbf{X}}
\def \bY{\mathbf{Y}}

\def \bphi{\mathbf{\Phi}}

\def \br {\mathbf{r}}

\def \bGamma {\mathbf{\Gamma}}

\def \be {\mathbf{e}}

\def \bN{\mathbf{N}}
\def \bY{\mathbf{Y}}

\newcommand \bu{\mathbf{u}}

\def \b1 {\mathbf{1}}

\begin{document}
\section{Introduction}
\vspace{-3mm}
Accurate seismic hydraulic fracturing monitoring (HFM) can mitigate many of the environmental impacts by providing a clear real-time image of where the fractures are occurring outside of the shale and how efficiently they are formed within the gas deposit. Although simple in principle, real time monitoring of hydraulic fracturing is extremely difficult to perform successfully due to high noise levels generated by the pumping equipment, anisotropic propagation of seismic waves through shale, and the multi-layered stratigraphy leading to complex seismic ray propagation, \cite{Eisner_TLE09,Aki_Book,Shearer_Book}. In addition the complexity of the source mechanism affects the relative amplitudes across the seismometers, \cite{Chapman_GJI012} introducing extra parameters in the system. Typical approaches for microseismic localization consists of de-noising of individual traces \cite{Liu_IGARSS09,Rodriguez_Geophy12} followed by time localization of the events of interest and then using a forward model under known stratigraphy to match the waveforms and arrival times, \cite{Oilfield_09,khadhraoui_SEG10}. The polarization estimation is achieved via Hodogram analysis \cite{Han_CREWES_2010_Thesis} or max-likelihood type estimation \cite{khadhraoui_SEG10}. In contrast to these approaches, recently the problem of moment tensor estimation and source localization was considered in \cite{Sacchi_GJI12} for general sources and in \cite{Ely_IGARSS12} for isotropic sources which exploit sparsity in the number of microseismic events in the volume to be monitored. This approach is shown to be more robust and can handle processing of multiple events at the same time.

Although our approach is very similar to the approach in \cite{Sacchi_GJI12} the main difference lies in the use of amplitude information from the Green's function. Here we don't use the amplitude (of the received waveform) information but only the \emph{temporal support information} or arrivals which is completely dictated by the velocity model of the stratigraphy and the source receiver configuration.  Since we are not using any amplitude information, we usually have more error in estimation and require more receivers for localization. Nevertheless, when the computation of Green's function is costly or accurate computation is not available our method can be employed. Furthermore, due to amplitude independent processing our methods can be extended to handle the anisotropic cases using just the travel-time information for inversion, \cite{Chapman_1992,Pratt_1992}.

\vspace{-3mm}
 \section{Microseismic source and data model}
 \vspace{-3mm}
 \label{sec:setup}
\begin{wrapfigure}[15]{l}{3.6 in}
\begin{center}
      \includegraphics[width = 3in, height = 1.75in]{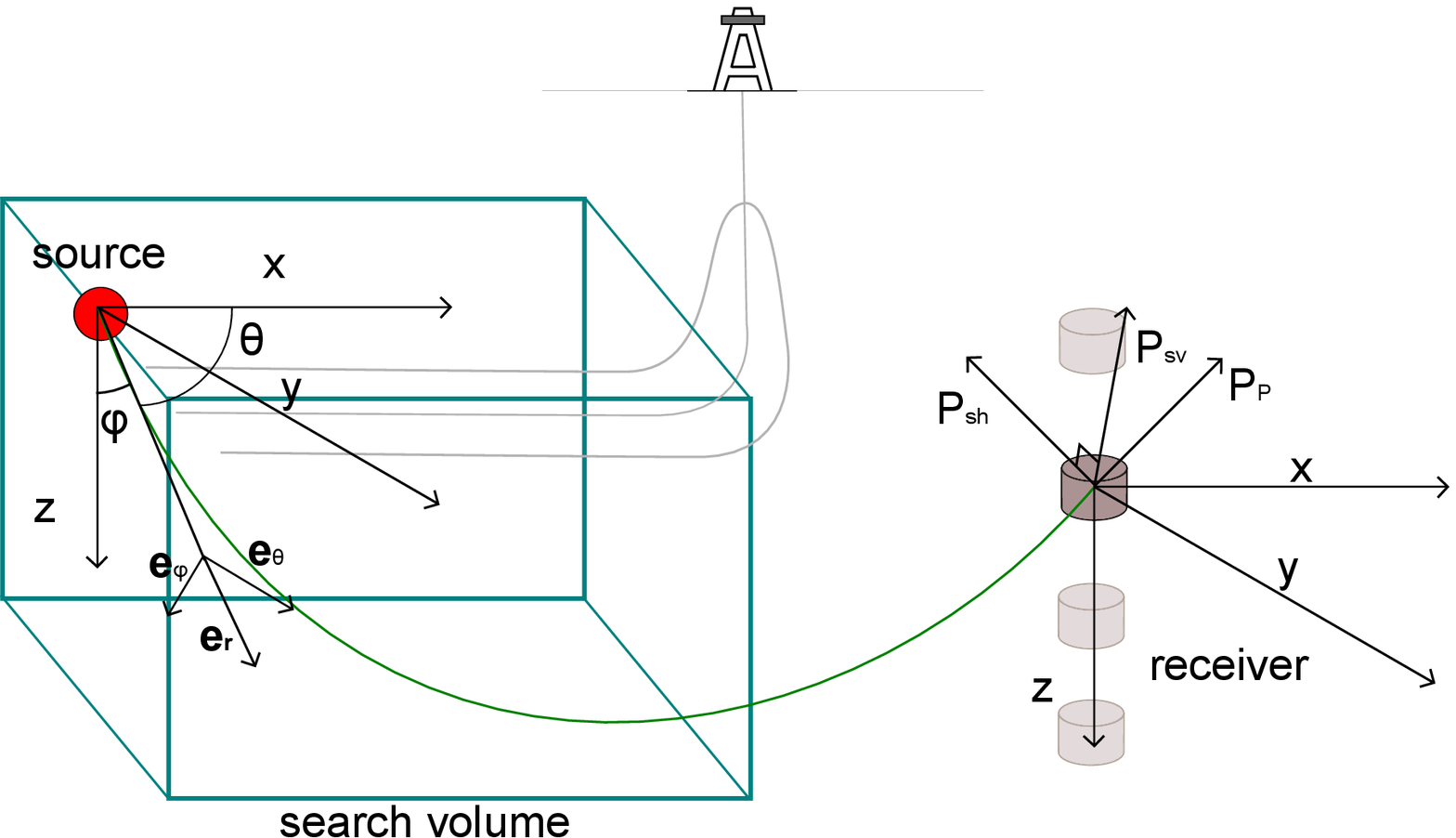}
      \end{center}
        \caption{This figure shows the geometry and coordinate system used in this paper.}
  \label{fig:setup}
\end{wrapfigure}In this paper we focus on isotropic layered media as the model for stratigraphy. The set-up is shown in Figure~\ref{fig:setup} where a seismic event with a symmetric moment tensor $\mathbf{M} \in \Real^{3\times3}$ is recorded at a set of $J$ tri-axial seismometers indexed as $j = 1,2,...,J$ with locations $\br_{j}$. Let the location of the source/seismic event be denoted by $l$. All these locations are with respect to a global co-ordinate system.  The seismometer records compressional wave denoted by p, and vertical and horizontal shear waves denoted by $sv$ and $sh$ respectively. Assuming (\cite{Aki_Book}, [Chapter 4]) that the volume changes over time does not change the geometry of the source, the particle motion magnitude vector (say) $\bu_c(l,j,t)$  at the three axes of the seismometer $j$ as a function of time $t$ , can then be described by the following equation,
\begin{align}
\bu_c(l,j,t) = \frac{R_c(\theta,\phi)}{4\pi d_{lj} \rho c^3} \,\,\mathbf{P}_{c}^{l j} \,\psi_c\left(t - \frac{d_{lj}}{v_c}\right)
\end{align}
where $d_{lj}$ is the radial distance from the source to receiver; $c \in \{p, sh, sv\}$ is the given wave type, and $\rho$ is the density, and $R_c$ is the radiation pattern which is a function of the moment tensor, the take off direction parameters $\theta_j, \phi_j$ with respect to the receiver $j$.  $\mathbf{P}_{c}^{l j}$ is the unit polarization vector for the wave $c$ at the receiver $j$. Up to a first order approximation \cite{madariaga_seismic_2007} we assume that $\psi_c(t) \approx \psi(t)$ for all the wave types and henceforth will be referred to as the source signal. Note that for isotropic formations and for compressional waves $\bP_{p}^{l j}$ is aligned with the incidence direction as determined by the ray propagation. The polarization vectors for the $sh$ and $sv$ correspond to the other mutually perpendicular directions.
The radiation pattern depends on the moment tensor $\mathbf{M}$ and is related to the take off direction at the source with respect to the receiver $j$ defined as the radial unit vector $\mathbf{e}_{r_j}$ relative to the source as determined by $(\theta_j , \phi_j)$, see Figure~\ref{fig:setup}. Likewise we denote by unit vectors $\be_{\theta_j}$ and $\be_{\phi_j}$ the radial coordinate system orthogonal to radial unit vector.
The radiation pattern for a compressional source $R_p(\theta_j,\phi_j)$ is then given by,
\begin{align}
R_p(\theta_j,\phi_j) = \mathbf{e}_{r_j}^T  \mathbf{M}  \mathbf{e}_{r_j} =
\left[
  \begin{array}{ccc}
    e_{r_{jx}}
    e_{r_{jy}}
    e_{r_{jz}}
  \end{array}
\right]
\left[
\begin{array}{ccc}
  M_{xx} & M_{xy} & M_{xz} \\
  M_{xy} & M_{yy} & M_{yz} \\
  M_{xz} & M_{yz} & M_{zz} \\
\end{array}
\right]
\left[
  \begin{array}{c}
    e_{r_{jx}} \\
    e_{r_{jy}} \\
    e_{r_{jz}}\\
  \end{array}
\right]
\end{align}
The radiation pattern can then be simplified and described as the inner product of the vectorized compressional unit vector product, $\textbf{e}_{p_j}$, and the vectorized moment tensor,
\begin{align}
R_p(\theta_j,\phi_j) =
\overbrace{\left[
  \begin{array}{cccccc}
    e_{r_{jx}}^2 & 2e_{r_{jx}} e_{r_{jy}} & 2e_{r_{jx}}e_{r_{jz}} & e_{r_{jy}}^2 & 2e_{r_{jy}}e_{r_{jz}} & e_{r_{jz}}^2 \\
  \end{array}
\right]}^{\textbf{e}_{p_j}^{T}} \mathbf{m}
\end{align}
where $\textbf{m} = \left[ M_{xx}, M_{xy}, M_{xz}, M_{yy}, M_{yz}, M_{zz}\right]^{T}$ and $(\cdot)^T$ denotes the transpose operation. The measurements regarding the moment tensor at the receivers can then be thought of as the measure of the corresponding radiation energy from the source. The above expression can then be used to construct a vector of radiation pattern $\mathbf{a}_p \in \Real^{J}$ across the $J$ revivers, with take off angles of ($\theta_j$ and $\phi_j$) corresponding to compressional unit vectors $\textbf{e}_{p_j}$, given by $\mathbf{a}_{p}= \mathbf{E}_p \mathbf{m}$ where $\mathbf{E}_p = [\be_{p_1},\be_{p_2},...,\be_{p_J}]^T$.
Similarly we have $ \mathbf{a}_{sh} = \mathbf{E}_{sh} \mathbf{m} $ and $ \mathbf{a}_{sv} = \mathbf{E}_{sv} \mathbf{m} $. Therefore we can write the radiation pattern across $J$ receivers for the three wave types as the product of an augmented matrix with the vectorized moment tensor.
\begin{align}
\label{eq:amp_model}
\mathbf{a} =
\left[
\begin{array}{l}
    \mathbf{a}_p  \\
    \mathbf{a}_{sh}   \\
    \mathbf{a}_{sv}  \\
  \end{array}
\right]
=
\underset{\mathbf{E}}{\underbrace{\left[
\begin{array}{l}
    \mathbf{E_p}  \\
    \mathbf{E_{sh}}   \\
    \mathbf{E_{sv}}  \\
  \end{array}
\right] }}\mathbf{m}
\end{align}
Thus the radiation pattern across the receivers $\textbf{a}$ can then be described as the product of the $\textbf{E}$ matrix, which is entirely dependent on the location of the event and the configuration of the array, and the vectorized moment tensor, which is entirely dependent on the geometry of the fault.

Under the above model for seismic source and wave propagation, given the noisy data at the tri-axial seismometers, the problem is to estimate the event location and the associated moment tensor. In contrast to existing work, our strategy for recovering the moment tensor consists of the following. First, we estimate the location of the source and the radiation pattern (vector) across the receivers using a sparsity penalized algorithm which is similar to the one used in \cite{Ely_IGARSS12} but modified to account for estimation of radiation pattern for non-isotropic sources. Following this we estimate the source signal and the radiation pattern using a singular value decomposition (SVD). The estimated radiation pattern is then used for the inversion for the moment tensor using the model given by Equation~(\ref{eq:amp_model}).

\vspace{-3mm}
\section{Formulation as a linear inverse problem}
\vspace{-3mm}
Our methodology rests on construction of a suitable representation of the data acquired at the receiver array under which seismic event can be \emph{compactly represented}. This compactness or sparsity in representation is then exploited for robust estimation of event location. We begin by outlining the following construction. \\
\begin{figure}[ht]
\centering \makebox[0in]{
    \begin{tabular}{c}
      \includegraphics[width = .6\textwidth]{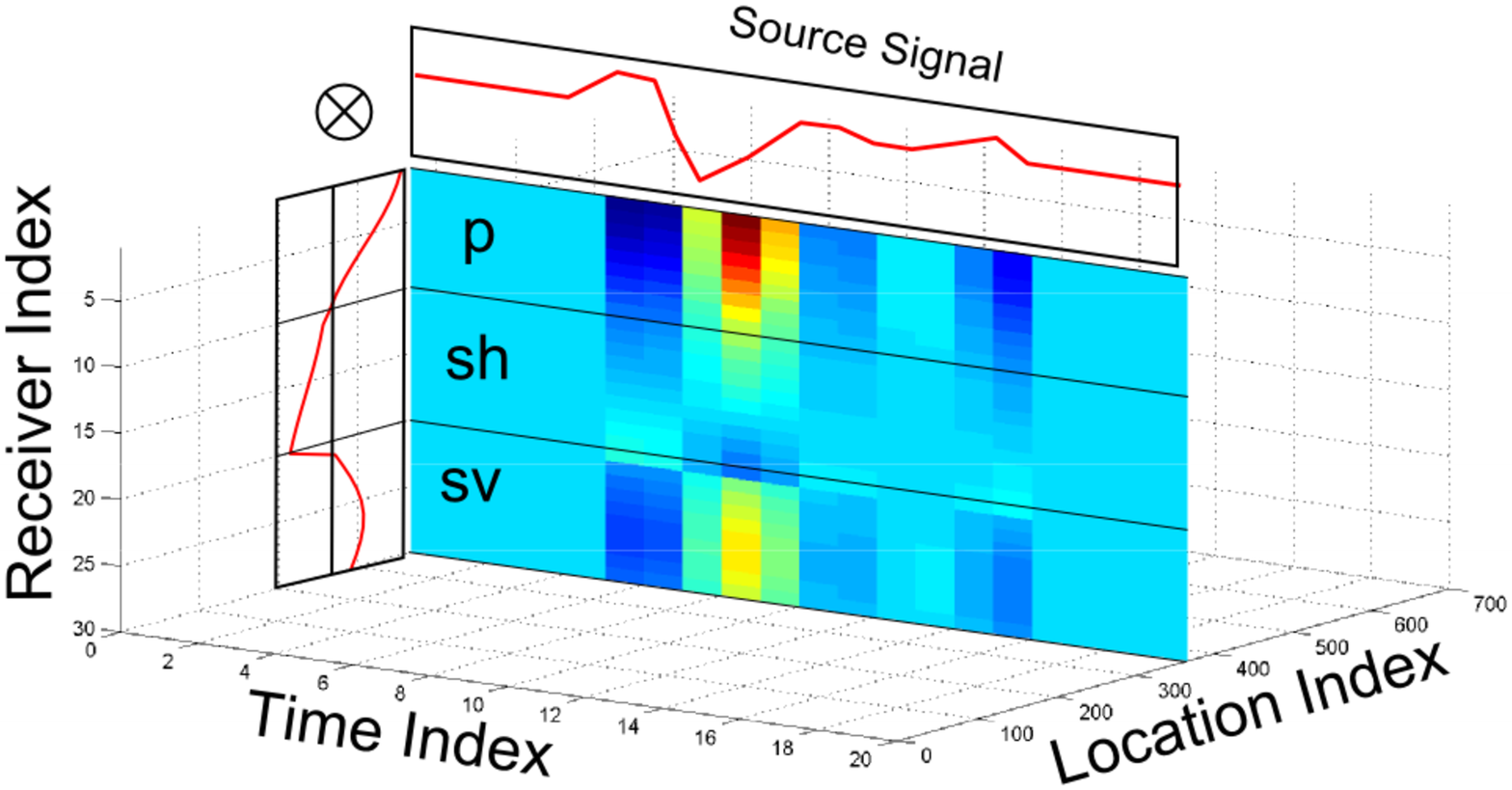}
      \includegraphics[width = .4\textwidth]{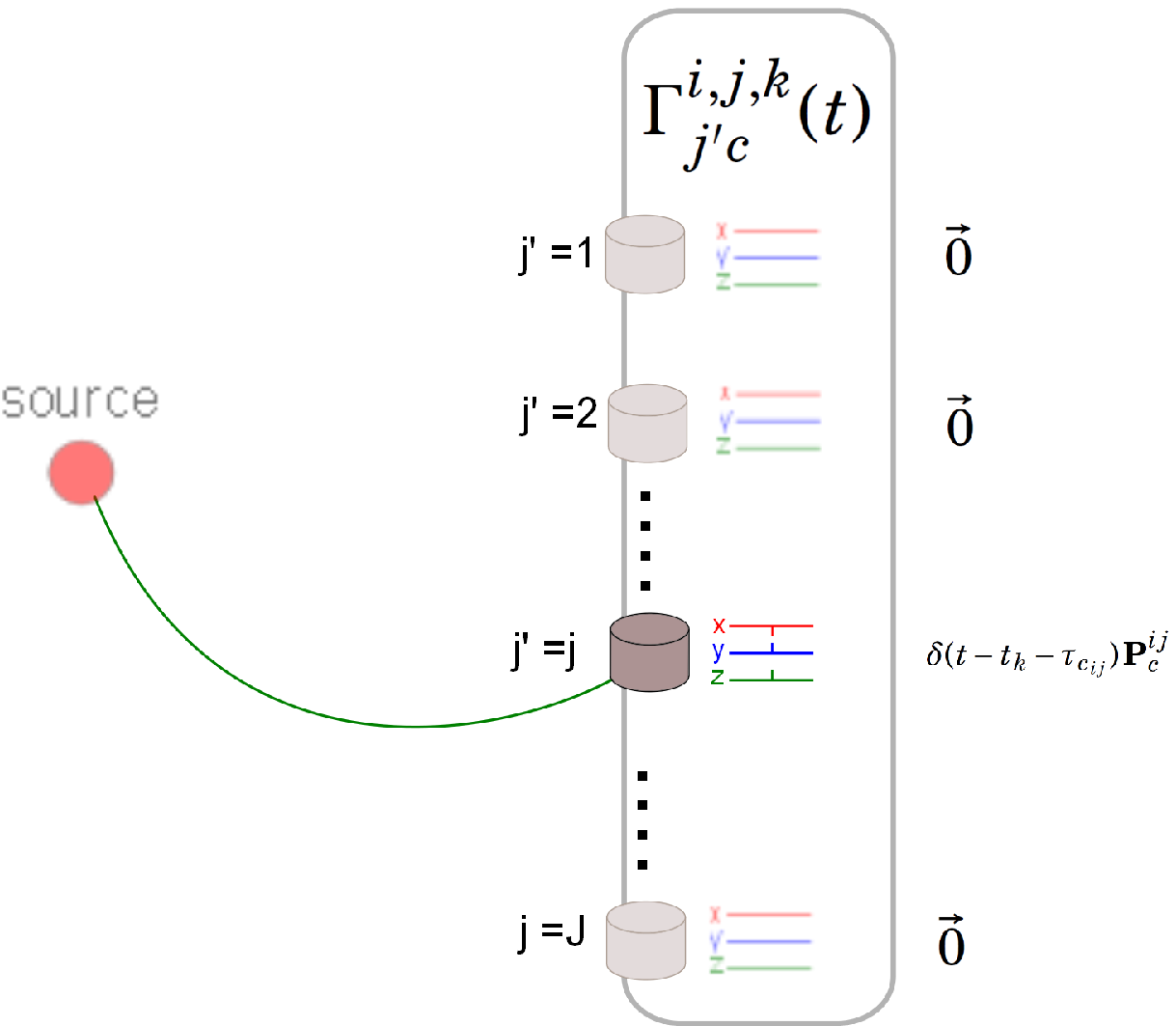}
      \end{tabular}}
  \caption{\textbf{Left:} This figure shows the block sparsity we exploit in our dictionary construction.  Note that the slice of the dictionary coefficients corresponding to the correct location of the event can be written as the outer product of the source signal and the amplitude pattern.   \textbf{Right:} This shows an example propagator.}
  \label{fig:sparsity}
\end{figure}
{\bf Representation of array data using space time propagators} - Assume that the source volume is discretized and the locations $l$ are indexed by $l = l_1, l_2,...,l_i,..,l_{n_V}$ where $n_V$ is the number of discretized locations. For a given location $l = l_i$ of the event, a fixed receiver $j \in \{1,2,...,J\}$ and wave type $c$, define $\bGamma_{c}^{i,j,k} = \{\bGamma_{j' c}^{i,j,k}(t)\}_{j' = 1}^{J}$, $t \in \mathbb{T}_r$, with $\mathbb{T}_r$ being the set of recording time samples at the receiver array, as the collection of the waveforms,
\begin{align}
\label{eq:propagator}
\mathbf{\Gamma}_{j' c}^{i,j,k}(t) =
\begin{cases}
        \delta(t- t_ k - \tau_{{c}_{ij}} ) \, \bP_{c}^{ij}  & \text{if } j' = j \\
        \vec{0} & \text{if } j' \neq j ,
\end{cases}
\end{align}
which corresponds to noiseless data at the \emph{single} receiver, $j$, as excited by an \emph{impulsive} hypothetical seismic event $i$ at location $l_i$ and time $t_k$ as shown in Figure \ref{fig:sparsity} right. Note that $\tau_{c_{ij}} = \frac{d_{l_i j}}{v_c}$ is the time delay and $\mathbf{\Gamma}_{j' c}^{i,j,k} \in \Real^{ |\mathbb{T}_{r}|\times J \times 3}$. For a given event location $l_i$ we collect these propagators to build
\begin{align}
\bGamma_{c}^{i,k} = [ \bGamma_{c}^{i,1,k}(:), \bGamma_{c}^{i,2,k}(:), \hdots, \bGamma_{c}^{i,J,k}(:) ] \in \Real^{3\,J\,|\mathbb{T}_{r}| \times J }
\end{align}
where $(:)$ denotes the MATLAB colon operator which vectorizes the given matrix starting with the first dimension. With this basic construction of the temporal and polarization response at the set of receivers for a given location $l_i$,   we construct a dictionary of propagators across the entire physical search volume indexed from $1$ to $n_{V}$, time support of the signal $t_k \in \mathbb{T}_s$,
\begin{align}
\Phi_c = \begin{bmatrix} \bGamma_{c}^{1,1}, \bGamma_{c}^{1,2}, \hdots, \bGamma_{c}^{i,k} ,\hdots, \bGamma_{c}^{n_V\cdot|\mathbb{T}_s|} \end{bmatrix}
\end{align}
We collect the overall dictionary of propagators for the three wave types into a single one,
\begin{align}
\Phi = \begin{bmatrix} \Phi_p \,\, \Phi_{sh}\,\,  \Phi_{sv} \end{bmatrix}
\end{align}
Clearly by construction and under assumption of superposition the data denoted as $\mathbf{Y} \in \Real^{|\mathbb{T}_r| \times 3\times J}$ can be written as $\bY(:) = \Phi \,\,\bX(:) + \bN$ where $\bN$ denotes the additive noise assumed to Gaussian, \cite{Liu_IGARSS09} and the \emph{coefficient} vector $\bX(:)$ is formed of the 3-D matrix $\bX \in \Real^{3\cdot J \times |\mathbb{T}_{s}| \times n_{V}}$ which captures the event location, excitation time of the source waveform $\psi(t)$ and the radiation pattern across the receivers for the three types of waves.

Under this modeling the problem is converted to estimation of $\bX$ from $\bY$ given (constructed) $\Phi$ which is a linear inverse problem. In presence of noise and under the severely ill-posed nature of the problem we will employ a regularized approach to inversion. In this context we note the following regarding the coefficient matrix $\bX$.

\vspace{-3mm}
\begin{enumerate}
\item Under the assumption that the number of primary seismic events per unit of time is small the matrix $\bX$ is \emph{sparse} along the third (location index) dimension, i.e. consists of a few non-zero \emph{frontal slices}.
\item  Each non-zero frontal slice (corresponding to the location of the event) is equal to $\boldsymbol{\psi} \,\mathbf{a}^{T}$ representing the amplitude (energy) variation of the source waveform across the receivers as a function of the moment tensor. This implies that each slice is a rank-1 matrix. This is illustrated in Figure~\ref{fig:sparsity} for a single event.
\end{enumerate}
\vspace{-3mm}

\section{Algorithm for location and moment tensor estimation}
\vspace{-3mm}
We now present an algorithmic workflow which systematically exploits these structural aspects for reliable and robust estimation of event location and moment tensor. The algorithmic workflow consists of three steps.\\

\vspace{-2mm}

\noindent {\bf Step 1: Sparsity penalized algorithm for location estimation} - Under the above formulation, we exploit the block-sparse, i.e. simultaneously sparse structure of $\bX$ for a high resolution localization of the micro-seismic events. The algorithm corresponds to the following mathematical optimization problem also known as group sparse penalization in the literature  \cite{Tropp2006SPa,Majumdar09}.
\begin{align}
\label{eq:minl12}
\hat{\bX} = \underset{\bX}{\arg\min}\,\,||\bY(:) - \bphi \bX(:)||_{2} + \lambda \sum_{i = 1}^{n_V} ||\bX(:,:,i)||_{2}
\end{align}
where $||\bX(:,:,i)||_{2}$ denotes the $\ell_2$ norm of the $i$-th slice, $\lambda$ is a sparse tuning factor that controls the group sparseness of $\bX$, i.e. the number of non-zero slices, versus the residual error.  The minimization operation was solved using the convex solver package TFOCS \cite{becker_templates_2010}. The parameter $\lambda$ is chosen depending on the noise level and the anticipated number of events. The location estimate  is then given by $l_{\hat{i}}$ where $\hat{i} = \underset{i}{\arg\max} ||\hat{\bX}(:,:,i)||_2$. In the following we denote the corresponding estimate of the $\hat{i}$-th slice $\bX(:,:,\hat{i})$ by $\hat{\bX}_{\hat{i}}$. \\

\vspace{-2mm}

\noindent {\bf Step 2: Estimation of waveform  and radiation pattern vector} - Once the optimization operation described in Equation~(\ref{eq:minl12}) is completed, the recovered slice, $\hat{\bX}_{\hat{i}}$, represents the source signal $\psi(t)$ modulated by the amplitude pattern across the receivers  and wave types, $\mathbf{a}$, i.e. $\hat{\bX}_{\hat{i}} = \boldsymbol{\psi} \mathbf{a}^{T}$. In order to estimate $\boldsymbol{\psi}$ and $\mathbf{a}$ we take the rank-1 SVD of $\hat{\bX}_{\hat{i}}$, where the right singular vector corresponds to the estimated source signal and the left singular vector to the estimated radiation pattern as shown in Figure \ref{fig:sparsity}.

{\bf Note}: The low-rank structure of the estimated matrix can then be used to detect if position and velocity model of the event were correctly estimated.  If the event is estimated correctly then the singular values of $\hat{\bX}_i$ should decay very rapidly.  If the decay is slow, then it is likely that the location is estimated incorrectly \footnote{Incorrect location estimates can result from poor resolution in discretization of the search volume or as a result of high degree of \emph{coherence} between neighboring location which are equally capable of explaining the data.}. We discuss some methods to deal with incorrect location estimates in Section~\ref{sec:sims}.

\noindent {\bf Step 3: Estimation of the moment tensor} -  Using the location estimate $l_{\hat{i}}$ and the knowledge of the source-receiver array configuration we construct the matrix $\mathbf{E}$ which is a function of $l_{\hat{i}}$ and the receiver configuration which is known and fixed. Then using the estimate of the radiation pattern $\hat{\mathbf{a}}$ from Step 2 we can write the simple inverse problem $\hat{\mathbf{a}} = \mathbf{E} \mathbf{m}$. However due to errors in estimation of $\mathbf{a}$ and ill-conditioning of $\mathbf{E}$ due to possible bad source-receiver configuration, one needs to again regularize for inversion. For this we use simple Tikhonov regularization approach where the moment tensor vector $\mathbf{m}$ is estimated via,
\begin{align}
\hat{\mathbf{m}}=((\mathbf{E}^T\mathbf{E}+\lambda_{m} \mathbf{I})^{-1}\mathbf{E}^T)\hat{\mathbf{a}}
\end{align}
where $\lambda_m$ is again tuned using some estimates on the uncertainty in estimation of $\mathbf{a}$ and according to the amount of ill-conditioning of $\mathbf{E}$.
\vspace{-3mm}
\section{Performance of the proposed algorithm on Synthetic Data}
\label{sec:sims}
\begin{figure}[htbp]
\centering \makebox[0in]{
    \begin{tabular}{c}
      \includegraphics[width = 1\textwidth, height = 3.2in]{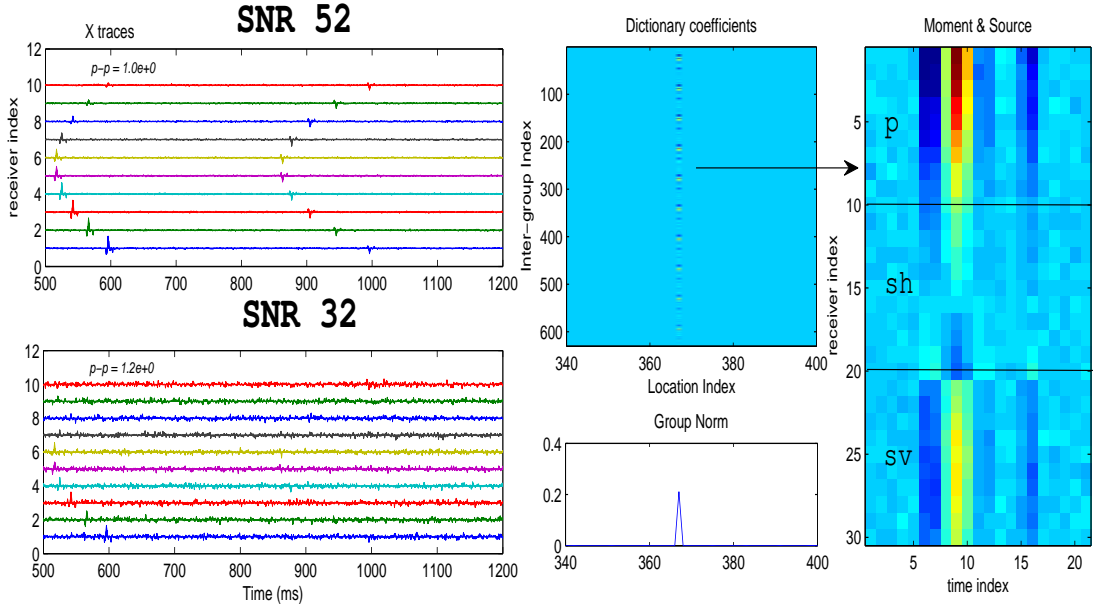}
      \end{tabular}}
  \caption{{\bf Left}: This figure shows the simulated traces along the x-axis for two noise levels. {\bf Top Middle}: This figure shows the value of the recovered dictionary coefficients.  {\bf Bottom Middle}:  This figure shows the vector of the $\ell_2$ norms of the slices of the coefficient matrix.  The largest value is taken as the location of the event. {\bf Right}:  This figure shows the dictionary coefficients of the corresponding location (slice) reshaped as a matrix.  Note that the source signal is common across the receivers and wave types.}
  \label{fig:results}
\end{figure}

\begin{figure}[h!]
\centering \makebox[0in]{
    \begin{tabular}{c}
      \includegraphics[width = 1\textwidth, height = 1.75in]{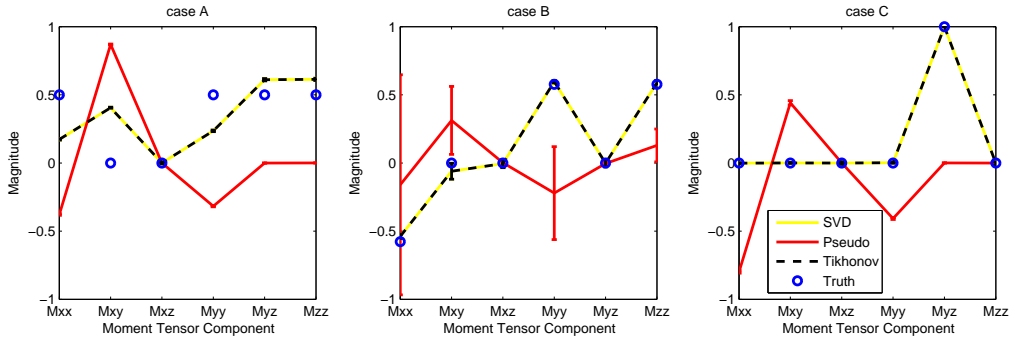}
      \end{tabular}}
  \caption{This figure shows the true moment tensor and estimated moment tensor using  the pseudo inverse with Tikhonov regularization and truncated SVD.  Without the regularization the estimates prove wildly inaccurate for all three of the test cases.  Regularization improves the estimated moment tensor in cases (b) and (c) but provides mixed results for test case (a).}
  \label{fig:momement}
\end{figure}

\vspace{-3mm}
{\bf Simulation set-up} - To test our proposed algorithm, synthetic data was generated for a single vertical well in a single layer isotropic medium with compressional velocity of 1500 m/s and shear velocity of 900 m/s.  Although an isotropic earth model is often unrealistic, we choose to use it in order to reduce the computational burden of a complex layered stratigraphy ray tracer.  It is clear that our approach does not take advantage of the isotropic model and can be easily extended to anisotropic and layered media without loss of generality.  The well is located at the origin with 10 sensors spaced 100 meters apart from a depth of 0 to 1000 meters.    For the first experiment a seismic event was simulated at (550,550,550) meters, in moderate noise resulting in an SNR of 46 dB, with three different moment tensors:  (a) isotropic mixed with shear slip,  (b)  compensated linear vector dipole mixed with isotropic and, (c) pure slip.  The true values of the simulated moment tensor are denoted by the blue dots in Figure~\ref{fig:momement}.  A 200 by 200 by 200 meter search volume was used with a spacing of 25 meters centered around the event.  The minimization operation in Equation~(\ref{eq:minl12}) was then used to determine the location of the event with the resulting localization by picking the slice with the largest $\ell_2$ norm shown in Figure~\ref{fig:results}.

\begin{figure}[h!]
\begin{center}
      \includegraphics[width = 3.6in, height = 2in]{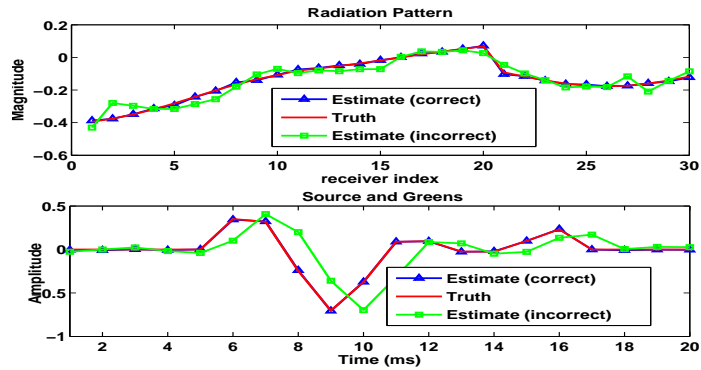}
      \end{center}
        \caption{{\bf Top}: This figure shows the true amplitude pattern and the estimated amplitude pattern.   {\bf Bottom}:  This figure shows the true and estimated source function.   For each of the two plots the reconstruction are shown for when the location of the event was estimated correctly and incorrectly.}
\label{fig:source_est}
\end{figure}

\noindent {\bf Radiation Pattern \&  Moment Tensor Recovery} - A rank 1 truncated SVD was then used to recover the source function and radiation pattern, as shown in figure \ref{fig:source_est}.   The estimated amplitude pattern was then inverted using a rank 5 truncated SVD and Tikhonov inversion with a $\lambda$ of $10^{-6}$, the ideal choice of truncation and $\lambda$ will vary as function of source receiver geometry and noise.
The simulation and estimation of the moment tensor and amplitude pattern was then repeated 20 times for each of the three cases.  For all instances the events were located at the correct location and the estimated moment tensors are shown in figure \ref{fig:momement}. Both Tikhonov and truncated SVD significantly resulted in nearly identical recovery of the moment tensor for cases (b) \& (c) and greatly improved the estimations of the non-regularized solution.  However, for test case (a) Tikhonov, SVD, and the non-regularized inverse provided poor estimates of the moment tensor.

 \begin{wrapfigure}[17]{l}{3 in}
\begin{center}
      \includegraphics[width = 2.7in, height = 1.75in]{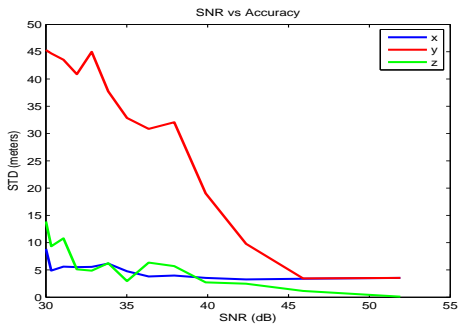}
      \end{center}
        \caption{This figure shows depth $z$, down range $x$, and cross range $y$ error as a function of SNR from 30 to 50 dB.  Both depth and down range estimations are much more robust to noise than the cross range estimate.}
  \label{fig:snr_loc}
\end{wrapfigure} 
 \noindent {\bf Location Accuracy} - In the second experiment we generated a seismic event with the moment tensor (a) with at the same location of (550,550,500) with an increased dictionary resolution of 5 meters. Gaussian noise was then added to the simulated trace for 20 noise levels with a resulting SNR of 25 to 50 and the location was estimated using equation \ref{eq:minl12} with a $\lambda$ of 3.  At each of the noise levels the process was repeated 20 times. Figure~\ref{fig:snr_loc} shows the resulting location accuracy (one standard deviation) as a function of SNR.  Depth $z$ and down range $x$ estimation proved to much more robust to noise than cross range estimations $y$.  The poor accuracy across the $y$ axes location is likely due to the fact that the estimation of the cross range of the event is highly dependent on the polarization amplitudes of the incident ray which is much more sensitive to noise than the estimate of arrival times. 

\vspace{-3mm}
\section{Conclusion \& Future Work}
\vspace{-3mm}
In this paper we have presented  comprehensive approach for HFM. The approach is robust towards uncertainties in stratigraphy models and is flexible to incorporate prior information at each step. For example, in Step 3 of the algorithmic framework one can use \emph{prior} information on $\mathbf{m}$ and make the inversion more robust. In this context we are currently looking to incorporate the distribution of eigenvalues of the matrix $\mathbf{M}$ \cite{baig_microseismic_2010} and exploit them in recovery of the moment tensor. Similar approach can be used in recovery of location estimate where prior information on location can be incorporated via a weighted penalty term like so $\sum_{i=1}^{n_V} w_{i} ||\bX(:,:,i)||_2$ in Equation~(\ref{eq:minl12}) where if $w_i$ is in inverse proportion to the likelihood of location $l_i$.

\begin{figure}[h!]
\centering \makebox[0in]{
    \begin{tabular}{c}
      \includegraphics[width = 1\textwidth, height = 2in]{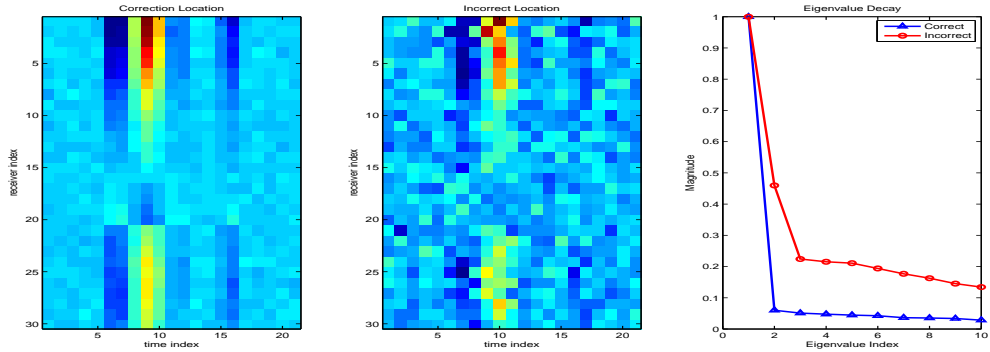}
      \end{tabular}}
  \caption{{\bf Left}: This shows the unwrapped dictionary slice for a seismic event when its location was estimated correctly.  {\bf Middle}: The unwrapped slice for an incorrectly located event.  Note that for the incorrect event the pattern of the source signal across the receivers is less constant and thus higher rank. {\bf Right}: this figure shows the normalized eigenvalues for the two matrices. For the correctly estimated matrix the eigenvalues decay rapidly. }
\label{fig:increase_rank}
\end{figure}

Estimation of the moment tensor proved difficult when the location of the event was estimated incorrectly.  When the location was estimated incorrectly the slice corresponding to the highest group norm could no longer be well approximated by a rank-1 outer product (figure \ref{fig:increase_rank}).  The resulting recovered radiation pattern and source function somewhat matched the simulated data but the source function was often time shifted and the radiation pattern was more noisy (figure \ref{fig:source_est}). Note that instead of a two step procedure to estimate the location followed by taking the SVD of the resulting estimates of the coefficient slices one can modify the algorithm of Equation~\ref{eq:minl12} to the following.
\begin{align}
\label{eq:minl1nuc}
\hat{\bX} = \underset{\bX}{\operatorname{\emph{argmin}}} ||\bY(:) - \bphi \bX(:)||_{2} + \lambda \sum_{i = 1}^{n_V} ||\bX(:,:,i)||_{*}
\end{align}
where $||\bX(:,:,i)||_{*}$ represents the nuclear norm of the $i$-th slice. In addition to implementing this proposed norm, we plan to validate our results using a more complex ray tracer on a anisotropic layered model.

\vspace{-3mm}
\bibliographystyle{jasanum}
\bibliography{ICA}
\end{document}